\documentclass[aps,prl,twocolumn]{revtex4}

\def\lesssim{\mathrel{\hbox{\rlap{\hbox{\lower4pt\hbox{$\sim$}}}\hbox{$<$}}}}
\def\gtrsim{\mathrel{\hbox{\rlap{\hbox{\lower4pt\hbox{$\sim$}}}\hbox{$>$}}}}
\def\geqq{\mathrel{\hbox{\rlap{\hbox{\lower4pt\hbox{$=$}}}\hbox{$>$}}}}

\usepackage{graphicx}

\begin{document}
\title {A model for giant flares in soft gamma repeaters}

\author{G. Lugones}\email{german.lugones@ufabc.edu.br}
\affiliation{Centro de Ci\^encias Naturais e Humanas, Universidade
Federal do ABC \\  Rua Santa Ad\'elia, 166 - Santo Andr\'e
09.210-170, SP - Brazil}

\author{E. M. de Gouveia Dal Pino}
\affiliation{Instituto de Astronomia, Geof\'\i sica e
Ciencias Atmosf\'ericas, Universidade de S\~ao Paulo\\
Rua do Mat\~ao 1226, 05508-900 S\~ao Paulo SP, Brazil}

\author{J. E. Horvath}
\affiliation{Instituto de Astronomia, Geof\'\i sica e
Ciencias Atmosf\'ericas, Universidade de S\~ao Paulo\\
Rua do Mat\~ao 1226, 05508-900 S\~ao Paulo SP, Brazil}


\begin{abstract}
We argue that giant flares in SGRs can be associated to the core
conversion of an isolated neutron star having a subcritical magnetic
field $\sim 10^{12}$ G and a fallback disk around it. We show that,
in a timescale of $\lesssim 10^5$ yrs, accretion from the fallback
disk can increase the mass of the central object up to the critical
mass for the conversion of the core of the star into quark matter. A
small fraction of the neutrino-antineutrino emission from the
just-converted quark-matter hot core annihilates into $e^+e^-$ pairs
above the neutron star surface originating the gamma emission of the
spike while the further cooling of the heated neutron star envelope originates the tail
of the burst. We show that several characteristics of the giant flare of
the SGR 1806-20 of 27 December 2004 (spike and tail energies,
timescales, and spectra) can be explained by this mechanism.
\end{abstract}


\maketitle

Soft $\gamma$-ray repeaters (SGRs) are persistent X-ray emitters
that sporadically emit short bursts of soft $\gamma$-rays. In the
quiescent state they have an X-ray luminosity of $\sim 10^{35}$
erg/s, while during the short $\gamma$-bursts they
release up to $10^{42}$ erg/s in episodes of about 0.1
s. Only four SGRs are known at present, 3 in our Galaxy, and one in
the Large Magellanic Cloud.  Exceptionally, three of them have
emitted very energetic giant flares which commenced with brief
$\gamma$-ray spikes of $\sim 0.2$ s, followed by tails lasting
hundreds of seconds. Hard spectra (up to 1 MeV) were
observed during the spike and the hard X-ray emission of the tail
gradually faded modulated at the neutron star (NS) rotation period.

The powerful giant flare of the SGR 1806-20 of  December 27, 2004 is
particularly important because it was observed by many satellites
\cite{Hurley2005,Mereghetti2005,Terasawa2005,Palmer2005,Schwartz2005}.
The duration of the initial spike was $\sim 0.25$ s. For the first
500 ms the emission was so strong as to saturate almost all the
detectors on $\gamma$-rays satellites, though there are unique
measurements of the initial flare activity \cite{Schwartz2005}. 
The luminosity of the spike was $\sim 10^{47} \textrm{erg/s}$ 
(assuming a distance of 15 kpc and isotropic radiation),
hundreds of times brighter than the giant
flares previously observed from other SGRs.  A long tail lasting
$\sim 400$ s followed the initial spike with a thermal component
with decreasing temperatures in the $\sim$ few  keV range
\cite{Hurley2005}. There is a clear modulation with the same period
of $7.56 \mathrm{s}$  already known from previous studies of the quiescent
emission. 

While several characteristics of SGRs are often explained in terms
of the \textit{magnetar} model, assuming that the object is a NS
with an unusually strong magnetic field ($B \sim 10^{15} $ G) \cite{woods}, 
there is alternative theoretical work
trying to explain the behavior of SGRs and Anomalous X-ray Pulsars (AXPs) by assuming the
existence of fallback disks around isolated NSs with subcritical
magnetic fields of $\sim 10^{12}$ G
(\cite{Chatterjee2000,Eksi_diagram,Eksi_hybrid,Ertan2007} and Refs. therein). NSs
are believed to be born from core$-$collapse supernovae with masses
near the Chandrasekhar limit of the Fe core ($\sim 1.4 M_{\odot}
$). During the supernova explosion, masses typically as large as
$\sim 0.2 M_{\odot}$ may fall back \cite{Fryer1999} and although
most of this material will be directly accreted onto the NS, part of
it can form an accretion disk within a few hours of the initial
explosion \cite{Chatterjee2000}. In fact, a fallback disk around an
AXP has been recently detected for the first time \cite{Wang2006}. In
this work we will show that accretion from a fallback disk can
increase the mass of the NS up to the critical value for the
conversion of the core into quark matter and this will trigger a
giant $\gamma$-flare.

\textit{The disk of SGRs:}
The inner disk radius $R_m$ of a fallback disk can be approximately
evaluated from the balance between the stellar magnetosphere stress
and the ram pressure of the inflow matter $R_m \sim
[\mu^2/(2GM)^{1/2} \dot{M}]^{2/7}$, where $M$ is the mass of the NS,
$\mu = R^3 B$ its magnetic moment, and $\dot{M}$ is the mass inflow
rate through the disk. If the NS magnetosphere rotates faster than
the inner disk, i.e., if $R_m > R_c$, where $R_c=
[GM/\Omega^2]^{1/3}$ is the corotation radius (defined as the radius
where the disk rotates with the angular velocity of the stellar
magnetosphere), then the system is in the so called
\textit{propeller} regime in which there is an angular momentum flux
from the NS to the disk (that will lead to strong stellar spin-down as
required by the observations of SGRs and AXPs).  In the case that the NS rotates only
a little faster than the inner disk ($R_m \gtrsim R_c$), a
significant fraction of the mass lost by the disk can accrete onto
the NS, while at the same time the NS remains spinning down in a ''tracking'' regime
\cite{Chatterjee2000,Ertan2007}. In recent work, Eksi et al.
\cite{Eksi_diagram} identified these different regimes in a stellar
period $P$ versus mass inflow rate $\dot{M}$ diagram.

As in \cite{Chatterjee2000,Eksi_diagram}, we shall assume here that
SGRs are NSs with moderate $B$-fields that spin down to $P
\approx 5-8$ s, as observed, in timescales $\lesssim 10^5$ yrs due
to a fallback disk. Notice that, if attributed to electron resonance, the
observed 5 keV cyclotron lines of 1806-20 \cite{Ibrahim} suggest
$B \sim 5 \times 10^{12}$ G. After a brief accretion phase that
follows its formation, the disk enters the propeller (spindown)
regime and later,  it reaches the tracking regime. During this
phase, the system will lie at the boundary between the propeller and
the X-rays luminous accretor (shaded) zone of the $P-\dot{M}$
diagram of Ref. \cite{Eksi_diagram}. According to that diagram, a NS with $B
\simeq 5 \times 10^{12}$ G can enter the tracking phase only if its
period is $\gtrsim 0.2$ s and $\dot{M} < 10^{20}$ g/s. Once reaching
this point, we assume that the disk will evolve along the tracking
path until the present observed period of 7.56 s (for SGR 1806-20)
which corresponds to a disk mass inflow rate  $\dot{M} \simeq 5 \times 10^{16}$
g/s \cite{Eksi_diagram}.

\textit{Quasi-steady luminosity and frequent soft $\gamma$-flares:} The
disk mass accretion rate onto the NS during the tracking
regime may consist of two contributions. One of them is a
quasi-steadily evolving sub-Eddington component, $\dot{M_{qs}}$,
whose value is of the order of ${\dot{M}}$ above. In the the
framework of the fallback disk model, the persistent X-ray
luminosity of $\sim 10^{35}$ erg/s produced during the more
quiescent states of the SGR 1806-20 is attributed to this component.
The maximum luminosity that a NS being presently powered by an
accretion rate $\dot{M_{qs}} \simeq \dot{M} \simeq 5 \times 10^{16}
g/s$ can produce is $L \simeq G M \dot{M}/R \, \simeq 2.7 \times
10^{36}$ erg/s, so that $4\%$ of the accretion power will be
sufficient to explain the observed persistent X-ray luminosity.
The other disk accretion rate component may be associated to the
observed soft $\gamma$-ray pulses of SGRs with luminosities $L \sim 10^{42}$
erg/s. A number of mechanisms have already been suggested to explain these
pulses. Among them, a hybrid model  (proposed, e.g., in
\cite{Eksi_hybrid}) assumes the existence of local
very strong multipole components of the  $B$-field with B $\sim
10^{14}-10^{15}$ G that act on the NS crust to produce the flares in
a similar way to the magnetar models, while the large scale dipole
component is $\sim 10^{12}-10^{13}$ G, as required in the fallback
disk scenario. Alternatively, we could speculate here that a sporadic super-Eddington
disk mass accretion rate ${\dot{M}}_{sp} \sim 5 \times 10^{21}$ g/s
onto the NS surface could produce the observed flares with $L \simeq
G M \dot{M}/R \sim 10^{42}$ erg/s (cf. \cite{mosquera}). Violent magnetic reconnection
events occurring between the NS magnetosphere and the inner disk
lines could trigger these events with the observed rise
times $\sim 10$ ms \cite{dalpino2005}.

\emph{Accretion and conversion of the core:}
The total accreted mass onto the NS during the source lifetime will
be the sum of both accreting
components above, and may be written as $\int{{\dot{M}}_{qs} } dt
 + \, N \tau_{sp} \times {\dot{M}}_{sp}$, where $N$ is the number of soft bursts during the
object's lifetime (several hundreds \cite{woods}) and the integral
is taken along the tracking path between $\dot{M}_{qs} \sim
10^{20}$ g/s and  $5 \times 10^{16}$ g/s. Using a self-similar evolution
for $\dot{M}_{qs} \propto t^{-\alpha}$, with $\alpha \sim 1.2$
\cite{Chatterjee2000}, we obtain a total accreted mass $\Delta M \sim 10^{-3} M_{\odot}$ in
a time interval $\lesssim 10^5$ yr, most of which comes from the
first component. This increase of the NS mass is essential
for triggering the conversion of the NS core into quark matter (QM).

In this paper we are {\it not assuming} the QM to be absolutely
stable (i.e. we assume that it has an energy per baryon at zero pressure and temperature
that is larger than the neutron mass).  Thus, stars containing QM are
\textit{hybrid}, i.e. contain $\beta$-stable QM only
at their interior and not up to the surface (as in strange stars).
The critical mass $M_{cr}$ for the  formation of a QM core in
a cold hadronic NS has been calculated recently exploring the effect of surface tension
and color superconductivity \cite{Bombaci2007}. If the
parameters of the hadronic and QM equations of state are such
that QM is \textit{not} absolutely stable, $M_{cr}$ is
very close to (but smaller than) the maximum mass of hadronic stars
for a large range of the parameter space \cite{Bombaci2007}. We emphasize that the conclusions of
the present paper are not sensitive to the exact value of $M_{cr}$
provided that it is larger than the NS mass short after the
supernova explosion and smaller than the maximum mass of NSs. Thus,
if right after the initial supernova fallback, the NS acquires a
mass that is not too far from $M_{cr}$, then the small accreted
$\Delta M$ above will be sufficient to trigger the conversion of the
core of the NS into QM at anytime after the beginning of the
tracking phase.

\emph{Energy release from the conversion:} The conversion into QM
of a large part of the core releases a large amount of energy
$\Delta E$ that can be estimated by the difference between the
gravitational mass of the initial and final configurations (both having
the same baryonic mass) \cite{Bombaci2007}. As shown recently \cite{Bombaci2007}, $\Delta E$
is $\sim 10^{53} - 10^{53.6}\mathrm{ergs}$ for the conversion of a hadronic
star having the critical mass. We assume that most of this energy is
thermalized inside the quark matter core, in the form of a trapped gas
of neutrinos and antineutrinos of all flavors ($\nu_i \bar\nu_i$ pairs).
To be sure, a small fraction can be used to excite vibrating modes \cite{guillerme},
but we shall neglect this mechanical energy in the rest of the paper.
Then, the initial temperature of the core
is estimated from $\Delta E \sim C_{q} T$, where $C_{q}$ is the heat capacity of the quark core, resulting
$T_{11} \simeq 1$ \footnote{We shall use the notation $Y_n = Y / 10^n[Y]$, where $n$ is
an integer, and $[Y]$ stands for the units of the quantity $Y$. We use cgs units and K for the temperature.}.

\emph{Neutrino annihilation:} Once they leave the opaque and hot
inner core, the  $\nu_i \bar\nu_i$ pairs can annihilate into
$e^+e^-$ pairs in the outer layers of the NS, and also above the
NS surface. Annihilation within the NS will heat matter
significantly while above the NS surface
it will create a $\gamma$-burst through $e^+e^- \rightarrow \gamma$.
For simplicity, we shall assume a blackbody spectrum for the $\nu {\bar{\nu}}$'s in the core.
Also, for annihilation above the NS surface, we can neglect
blocking effects in the phase spaces of $e^-$ and $e^+$.
Thus, the "unblocked" local energy deposition rate at a radial position $r$ due to
$\nu_i \bar\nu_i \rightarrow e^+e^-$ can be written as $Q^{unb}_{e^+ e^-}
= \mathcal{A} T_{\nu11}^9 \chi[x]$, being $T_{\nu11}$ the temperature
at the $\nu$-sphere  (i.e. the last scattering surface of the $\nu {\bar{\nu}}$'s),
$\mathcal{A} = 1.28 \times 10^{34} \mathrm{erg ~ cm^{-3}s^{-1}}$
for $\nu_e \bar\nu_e$'s, and $\mathcal{A} = 2.7 \times 10^{33} \mathrm{erg ~ cm^{-3}s^{-1}}$
for $\nu_{\mu} \bar\nu_{\mu}$'s and $\nu_{\tau} \bar\nu_{\tau}$'s \cite{annihilation}.
The angular factor $\chi[x] = \frac{1}{8} (1 - x)^4 (5 + 4 x + x^2)$
depends on $x(r) = (1 - R^2_{\nu} / r^2 )^{1/2}$, being $R_{\nu}$
the $\nu$-sphere radius. $\chi[x]$ is responsible for the rapid
decrease in $Q^{unb}_{e^+ e^-}$ for increasing $r$, as the interacting
neutrinos become more collinear.
The total luminosity injected \textit{above} the NS surface is
determined by $L_{e^+ e^-} = \int_R^{\infty} Q^{unb}_{e^+ e^-} 4 \pi r^2
dr = \xi \mathcal{A} T_{\nu}^9 R_Q^3$
where $\xi = \int_{x(R)}^{1} 4 \pi x  (1
- x^2)^{-5/2} \chi[x] dx$ is a function of the ratio between $R_{\nu}$
(which we assume to be coincident with the quark-matter core
radius $R_Q$) and the NS radius $R$.
Within the NS, electrons are degenerate and there is a strong
blocking effect over $Q_{e^+ e^-}$. This can be accounted for by applying
an average blocking factor to the unblocked results \cite{annihilation}. Then, one writes
$Q_{e^+ e^-} = B_e  Q^{unb}_{e^+ e^-}$, where
$B_e \sim 1/[ 1 + \exp((\mu_e - T_{\nu})/T)]$, $\mu_e$ is
the $e^-$ chemical potential at the annihilation point, and $T$ the corresponding temperature \cite{annihilation}.

\emph{Cooling evolution of the NS:} The conversion of the inner core
region into quark matter is expected to be very fast
($\sim 10^{-5}$ s \cite{Lugones1994}). We assume for simplicity that after the conversion,
the  NS settles reasonably fast into a nearly hydrostatic configuration.
Though this assumption is not strictly correct, particularly during the
first seconds of the evolution, we shall see below that the characteristics of the
$\gamma$-ray tail over 400 s are determined by a thin shell
near the surface of the NS which is much less sensitive to dynamical
rearrangement than the core region. As we will also see, it is the neutrinos from
the core that produce the spike. Even if the core is subject
to dynamical variations these will hardly affect the order of magnitude
of the released energy and the $\nu$-emission timescale in comparison to the hydrostatic case (cf. \cite{guillerme}).
Therefore, soon after the conversion we adopt a simplified model in which
the NS (with $M = M_{cr}$) has a hot quark matter inner-core with $R_Q \approx 0.6 R$
in equilibrium with a trapped $\nu {\bar{\nu}}$ gas at $T_{\nu11} \approx 1$ surrounded
by an \textit{initially cold} outer-core (composed of free nucleons, electrons and muons) and  a crust
(composed of nuclei, free neutrons and electrons).  We assume there is no significant
lepton number gradient, since most of the $\nu$'s are produced in pairs, so that the
lepton-number-transport equations will not dominate the evolution.
Thus the cooling of the quark core is described by the energy-transport equation:
\begin{equation}
c_q \frac{\partial T}{\partial t}  =
  \frac{\Gamma}{r^2 e^{\Phi}} \frac{\partial}{\partial r}
 \bigg[ e^{2\Phi} r^2  \frac{c \lambda}{3} \frac{\partial \epsilon_{\nu}}{\partial r}
 \bigg]
\end{equation}
\noindent where the neutrino energy density is  $\epsilon_{\nu} = 3 \times \frac{7}{8} a T^4 $, 
$a = \pi^2 k_B^4 /15 c^3 \hbar^3$,  and $c_q$ is the specific heat of quark matter.
The neutrino mean free path $\lambda \approx 1.3 \times 10^3 \mathrm{cm} / T_{11}^3$
in the hot inner-core is very small and the $\nu {\bar{\nu}}$'s  are released
in a diffusion timescale, due to scattering with free quarks. At the surface of the quark core we assume a free
streaming regime into vacuum. Solving numerically this equation we find the time evolution
of the temperature $T_{\nu}$ at the $\nu$-sphere.  $T_{\nu}$ is then used to evaluate the 
luminosities   $L_{e^+ e^-}$ given above and $L_{\nu} = 4 \pi R_Q^2 \times 3 \, \frac{7}{8} \sigma_{SB} T_{\nu}^4$ (Fig. 1).

\begin{figure}
\includegraphics[angle=-90,width=8.5cm]{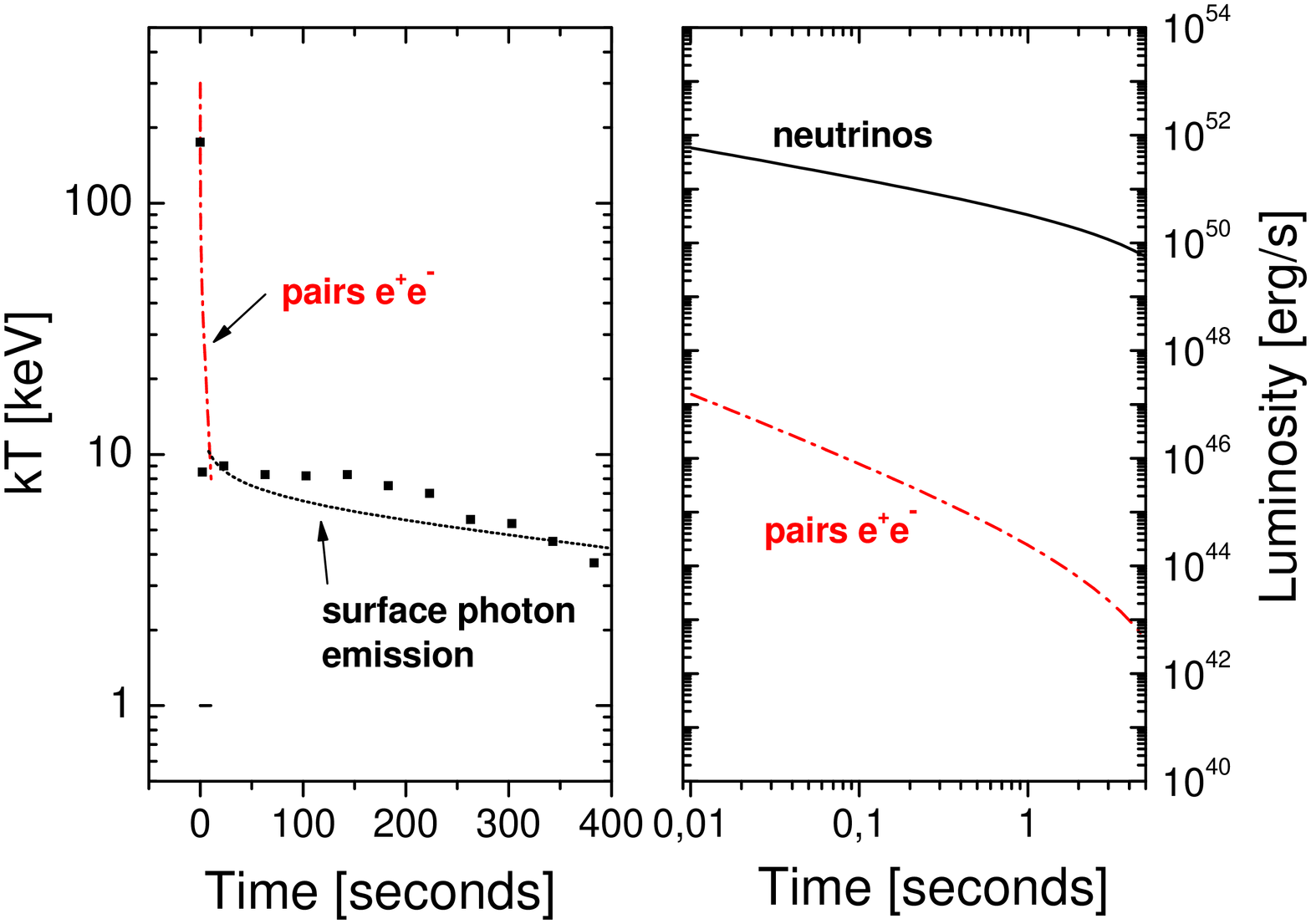}
\caption{\textit{Left:} The emission spectrum
according to the present model (dashed line) compared with the
observed spectrum of SGR 1806-20 (dots with errors included) given
in \cite{Hurley2005}. \textit{Right:} Luminosity of the core neutrino emission
and of the  $e^+e^-$ injection above the NS surface ($\approx L_{\gamma}$ of the spike) 
during the first few seconds after the conversion.  }
\end{figure}

The cold outer-core and the crust are essentially transparent to $\nu {\bar{\nu}}$'s
and therefore they will hardly heat due to the emission of the hot inner-core.
Also, since  $e^-$'s are strongly degenerate, heating of a given layer through
$\nu \bar\nu \rightarrow e^+e^-$ will be significant only
if the mean energy of the $\nu {\bar{\nu}}$'s (which is of the order of that of the outgoing $e^+e^-$)
is larger than the Fermi energy of the  $e^-$'s at that layer, i.e. $k_B T_{\nu} \gtrsim \mu_e$.
Since $T_{\nu} \lesssim 10$ MeV, this is fulfilled only in the outer-crust/envelope. Using a simple relation
between the depth $z$ measured downward from the NS surface and $\mu_e$ \cite{hernquist},
$z \approx  \mathrm{25 m} (\mu_e/m_e - 1)
\approx \mathrm{25 m} [T_{\nu} / m_e - 1] $, we find that initially only the outermost
$\sim 500$ meters will be significantly heated by this process.
Since the calculations show that $T_{\nu}(t)$ decreases exponentially (Fig. 1), heating
ceases in less than $\sim 1$ s and thereafter, the envelope evolves
decoupled from the core. The cooling of the envelope
can then be described in a plane parallel approximation by:
\begin{equation}
\frac{\partial T}{\partial t}  =
  \frac{\Gamma e^{\Phi}}{ c_v} \frac{\partial}{\partial z}
 \bigg[ K \frac{\partial T}{\partial z}
 \bigg] - e^{\Phi} Q_{\nu} \,   ,
\end{equation}
where the specific heat is approximately that of a Dulong-Petit solid and a degenerate relativistic electron gas, $c_v \approx (4.5 \times 10^{16} \rho_{10} + 3.9 \times 10^{16} \rho_{10}^{2/3} T_9 ) \, \mathrm{erg \, cm^{-3} K^{-1}}$. The thermal conductivity is $K = 16 \sigma_{SB} T^3 / (3 \rho \kappa)$ with $\kappa^{-1} = \kappa_{cond}^{-1} + \kappa_{rad}^{-1}$  where the conductive opacity is $\kappa_{cond} = 3.6 \times 10^{-2} \rho_6^{-1.2} T_9^{1.4} \mathrm{cm^2 g^{-1}}$  and the radiative opacity is $\kappa_{rad} = 9.6 \times 10^{-2} \rho_6^{0.1} T_9^{-1.2} \mathrm{cm^2 g^{-1}}$ (within a factor of $\sim 3$) \cite{eichler2002}. The density profile is given by $\rho_6 \sim (z /10^4 \mathrm{cm})^3$ \cite{hernquist} and the effective surface is taken where the optical depth is $\sim 2/3$. For simplicity,  the  $\nu$-cooling sink is given approximately by $Q_{\nu} = 1.03 \times 10^{15} T_9^{9.6} \mathrm{erg \, cm^{-3} s^{-1}}$ for $T_9 > 1.3$ and $10^{-2} \lesssim \rho_6 \lesssim 10^{-2}$ \cite{eichler2002}. The energy per unit volume injected by core-$\nu$'s in the envelope is  $U(z) \sim \int B_e \mathcal{A} T_{\nu11}^9(t) \chi[x] dt$. Assuming that this energy is thermalized the initial temperature profile is given by $\sim U(z)/c_v(z) > 10^{10}$ K, but due to the efficient $\nu$-cooling sink it falls to values  $\lesssim 10^{10}$ K in less than 1 s. Thus the later evolution of the envelope is rather insensitive to the initial energy injection.

\emph{Results:} In the right panel of Fig. 1, we show
the time evolution of the core neutrino luminosity and the luminosity
$L_{e^+ e^-}$ of the pairs  that annihilate above the NS's surface. The
timescale at which  $L_{e^+ e^-}$  decays to $\sim 1 \%$ of its initial value is
$\sim 0.2$ s, just as observed for the spike of giant flares in SGRs (this is a robust
result that naturally arises also in models of GRBs involving the burning of
quark stars; e.g. \cite{Haensel1991}). We also note that
the efficiency of the $\nu {\bar{\nu}} \rightarrow e^{+}e^{-}$ conversion is strongly
temperature dependent, so that with variations within
a factor of $\sim 2-3$ in the central temperature of the object, a wide range of observed spike-luminosities
($10^{44}-10^{47}$ erg/s) can be explained.
The spectrum of the spike has been calculated assuming that all pairs
are converted into $\gamma$-rays (i.e. $ L_{e^+ e^-} \approx L_{\gamma}$). This results $T_{\gamma}(t) \propto T_{\nu 11}^{9/4}(t)$ which reproduces the observed best-fit blackbody temperature $T_{spike} = 175 \pm 25$ keV of SGR 1806-20 \cite{Hurley2005} for an initial core temperature $T_{\nu 11} \approx 0.5$ (left panel of Fig. 1).
The luminosity $L_{e^+ e^-}$ fades very fast and then is
overwhelmed by photon radiation from the hot NS envelope. The surface
temperature evolves according to Eq. (2) producing the tail emission as shown in Fig. 1.

\emph{Discussion:} The model just described can
explain the energy scale, spectrum and timescale of both the spike
and the tail of giant flares in SGRs.  Several features can be used
to observationally distinguish between this and the magnetar model.
For example, the concomitant detection of low energy neutrinos and
anti-neutrinos ($E_{\nu} \approx 10$  MeV) with the
$\gamma$-emission will give strong support to our model. Unlike in
the magnetar model, a huge amount of energy stored in the core is
converted in $\gamma$-emission with just a tiny efficiency. The
eventual observation of a giant flare with energy much larger than
$10^{47}$ erg will also give support to this model, while it would
be hard to explain within the magnetar scenario. Also, all four SGRs
are possibly associated with supernova remnants (SGR 1900+14 /
G42.8+0.6, SGR 0526-66 / N49,  SGR 1806-20 / G10.0-0.3, and SGR
1627-41 / G337.0-0.1) which makes it plausible that they still have
a fallback disk. The recent identification of a fallback disk around
the AXP 4U 0142+61  \cite{Wang2006} points to this direction,
provided that SGRs and AXPs belong to the same population. The
release of energy in the core and the NS-structure rearrangement
after phase transition, as well as the sporadic impact of accreted
matter, may cause surface stresses and quakes just as those that in
the magnetar scenario are caused by magnetic stresses. Therefore,
quasi-periodic-oscillations (QPOs), for example, could be
likewise explained by crust fracturing and release of elastic
energy. Finally, since AXPs seem to have bursts but not giant
flares, we can interpret them as accreting NSs whose mass is far from $M_{cr}$. In this sense, the
number of AXPs should be larger than that of SGRs. Also, our model
predicts that all SGRs that had a giant flare should have
essentially the same mass, i.e. $M_{cr}$. Techniques for
determination of NS-masses are rapidly improving and eventual
measurements could probe the value of $M_{cr}$. If so,
this could also provide key information for the equation of state of
dense matter.

\acknowledgements{The authors acknowledge partial support from FAPESP and CNPq.}

\end{document}